\documentclass[11pt,twoside]{article}
\usepackage{asp2010}
\usepackage[pdfauthor={A. Accomazzi},pdftitle={ADS: The Next Generation Search Platform}]{hyperref}

\resetcounters
\aspvolume{TBD}
\aspcpryear{2014} 
\aspvoltitle{Library and Information Services in Astronomy VII} 
\aspvolauthor{Andras Holl, Soizick Lesteven, Dianne Dietrich, and Antonella Gasperini, eds.} 

\begin{document}
\title{ADS: The Next Generation Search Platform}
\author{Alberto Accomazzi, %\orcid{0000-0002-4110-3511}
Michael J. Kurtz, 
Edwin A. Henneken, 
Roman Chyla, 
James Luker, 
Carolyn S. Grant, 
Donna M. Thompson, 
Alexandra Holachek, 
Rahul Dave,
Stephen S. Murray
}
\affil{Harvard-Smithsonian Center for Astrophysics, 60 Garden Street,
  Cambridge, MA 02138, USA}

\begin{abstract}
Four years after the last LISA meeting, the NASA Astrophysics Data System (ADS) 
finds itself in the middle of major changes to the 
infrastructure and contents of its database.  In this paper we highlight a number of features of 
great importance to librarians and discuss the additional functionality that we are currently developing.
Starting in 2011, the ADS started to systematically collect, parse and index full-text documents 
for all the major publications in Physics and Astronomy as well as many smaller Astronomy journals and 
arXiv e-prints, for a total of over 3.5 million papers.  Our citation coverage has doubled since 2010 and 
now consists of over 70 million citations.  We are normalizing the affiliation information in our records and, 
in collaboration with the CfA library and NASA, we have started collecting and linking funding sources with 
papers in our system.
At the same time, we are undergoing major technology changes in the ADS platform which affect all aspects 
of the system and its operations.  We have rolled out and are now enhancing a new high-performance search 
engine capable of performing full-text as well as metadata searches using an intuitive query language which 
supports fielded, unfielded and functional searches.  We are currently able to index acknowledgments, 
affiliations, citations, funding sources, and to the extent that these metadata are available to us they 
are now searchable under our new platform.  The ADS private library system is being enhanced to support 
reading groups, collaborative editing of lists of papers, tagging, and a variety of privacy settings when 
managing one's paper collection.
While this effort is still ongoing, some of its benefits are already available through the ADS Labs user 
interface and API at \url{http://adslabs.org/adsabs/}.

\end{abstract}

\section{Introduction}

The ADS was originally conceived over 20 years ago as a system to support the discovery and retrieval 
of data from the NASA Astrophysics missions and the scholarly literature written about it \citep{2000A&AS..143...41K}.
With the restructuring of the ADS program in 1994, the system became the primary service providing 
a bibliographic discovery platform to researchers in Astronomy, Astrophysics, and related fields.  
Today, the ADS is best described as a ``disciplinary repository'' for bibliographic content in
Astronomy and Physics.  In addition to its search capabilities, ADS tracks 
citations between papers, links to datasets associated with the publications, provides article-level metrics,
and features personalized notification services to registered users.

Over its lifetime, ADS has seen an astonishing growth in its data holdings and capabilities.
The current number of bibliographic records in ADS is now above 10 million and covers all of the refereed
literature in Astronomy and Physics, plus all of the so-called ``gray literature'' of interest to 
astronomers.  This consists of conference proceedings, meeting abstracts, PhD theses, technical reports,
observing proposals, descriptions of data catalogs hosted by Vizier \citep{2000A&AS..143...23O}, 
and software packages used in astronomy research \citep{2013ASPC..475..387A}.
The ADS citation database now consists of over 5 million nodes (papers with references or citations in the ADS 
database) connected by over 72 million edges (links in the network between cited and citing papers).

It is important to point out that ADS is not just your prototypical bibliographic database, but is
better seen as an aggregator of scholarly resources.  Rather than simply index content from a variety of
publishers, the ADS provides a number of value-added services which enrich its collection of records
in a number of different ways.  We harvest and merge bibliographic data from multiple sources
(arXiv, CrossRef, a number of publishers, and Astrophysics archives), cross-correlating content across
them (arXiv e-prints and published papers, translations, re-publications, Vizier catalogs, observing and
funding proposals and their papers).  We collect and expose links to external resources such as 
publishers and data archives (SIMBAD, NED, Vizier, other NASA and ESA data centers, etc.).  
We enrich article metadata via text-mining of full-text sources (extracting references, acknowledgments,
keywords).  We incorporate bibliographies from institutes and archives.  And we generate and maintain 
citation and co-readership networks \citep{2007ASPC..377...69A,2005JASIS..56..111K}.

The ADS user base has been steadily expanding during the past two decades and includes every working
astronomer, a number of researchers in physics, librarians, archivists, and the general public.
Our log analysis shows that there are now over 55,000 heavy users of our system (these are people
who access the ADS on a weekly basis and account for more than 10 visits per month).  There are over 1 million
infrequent ADS users on any given month.  Part of the reason for this large user base is due to the fact
that metadata in ADS is indexed by the main search engines (Google, Bing, Yahoo) and that Wikipedia has
over 80,000 links to ADS pages.  Thus a service that was developed by Astronomers for Astronomers has now
become a reference digital library easily accessed by scholars and the general public alike.

In this paper we provide an update on recent ADS developments, including the technical infrastructure
being implemented, the functionality supported by the new platform, and features being rolled out
in the near future.

%%%%%%%%%%%%%

\section{The New ADS Platform}

In 2009 the ADS team started working on a major overhaul of its technical infrastructure in order to
replace its aging, home-grown search engine and bibliographic database with state of the art
technology.  The main motivation behind this effort was the creation of a robust
technical infrastructure built on modern web and digital library standards which could support
current and planned ADS functionality.  Here we present some of the major changes in the
ADS platform, \href{http://adslabs.org/adsabs}{as currently deployed under ADS Labs},
highlighting the features of interest to librarians and power-users.

\subsection{Search Engine}

The search engine we are working on
is a custom-built version of the open-source \href{http://lucene.apache.org/solr/}{Apache SOLR search server},
heavily modified to efficiently index and search bibliographic metadata, full-text contents,
citation and usage data.
SOLR is an enterprise search server built on the Apache Lucene 
library, which provides a state-of-the-art implementation of search engine capabilities.
While ADS ``Classic'' was built from the ground up using the restricted set of tools available
to us in the early 1990s, we believe that the adoption of a system based on a widely used open-source 
search engine will provide us with a durable platform for some time to come, reaping the benefits
of the open source model.

The new ADS search engine inherits all the benefits of Lucene, which features
an advanced parser supporting queries on single terms as well as phrases, 
fielded and unfielded searches, wildcard, proximity, and fuzzy searches, 
synonym and acronym expansion, boolean operators and groupings.
This search syntax has been extended by the ADS team to support functional operators, which provide
a way to implement advanced capabilities by taking an input query and manipulating its results.  
The operators currently implemented fall into the three categories below.

\begin{itemize}
\item {\bf Post-processing operators.}
Two functions provide a way to limit search results by applying a post-processing
filter to the list of document selected by the original query. The {\it pos()} operator
allows one to search for an item within a field by specifying the position of the item within the field. 
The syntax for this operator is {\it pos(fieldedquery,position)}.
The {\it topn()} operator allows one to limit the number of results returned by the search engine.
The syntax for this operator is {\it topn(N,query,[ranking])} and returns the top N papers
from a list of ranked results; this set is often useful as an input list for further queries.

\item {\bf Citation-based operators.}
Two functions are provided to allow the user to retrieve the list of references or citations
generated from an original list of results.  The {\it references(query)} operator returns the list
of papers referenced by the documents selected by the input query.  Similarly, the 
{\it citations(query)} operator returns the list of papers citing the documents selected by
the input query.

\item {\bf Second-order operators.}
In order to support a variety of discovery modes, 
three additional operators have been defined based on the algorithms described in
\citep{2002SPIE.4847..238K}.  Each of them takes an input query about a 
particular topic, computes a list of results, and returns papers related to them 
according to different criteria.
The {\it instructive()} operator returns the list of documents citing 
the most relevant papers on the topic being researched; the operator is named ``instructive''
because it selects papers containing the most extensive reviews in this field.
The {\it useful()} operator returns the list of documents frequently cited by the most 
relevant papers on the topic being researched; the operator is named ``useful''
because it selects papers discussing methods and techniques useful to conduct research in this field.
The {\it trending()} operator returns the list of documents most read by users who read 
recent papers on the topic being researched; the operator is named ``trending'' 
because it selects papers currently being read by people interested in this field.
\end{itemize}

A sample of the queries supported by the new ADS search engine helps illustrate the
functionality currently implemented (see table 1).

\begin{table}[!ht]
\caption{A sample of queries supported by the new ADS search engine.
For a full description of the syntax, please 
\href{http://adslabs.org/adsabs/page/help/search}{see the online help}.}
\smallskip
\begin{center}
{\small
\begin{tabular}{ll}
\tableline
\noalign{\smallskip}
Query type & Sample query \\
\noalign{\smallskip}
\tableline
\noalign{\smallskip}
Search by author & {\it author:``Accomazzi, Alberto''} \\
First author only & {\it author:``\^{}Accomazzi, Alberto''} \\
Abstract search & {\it abs:(gravitational lensing)} \\
Abstract phrase search & {\it abs:``gravitational lensing''} \\
Search the full text & {\it full:(HST or JWST)} \\
Acknowledgments & {\it ack:``ESO''} \\
Unfielded search (AND) & {\it Accomazzi ADS bibliography} \\
Unfielded search (OR) & {\it Accomazzi or ADS or bibliography} \\
Search by affiliation & {\it aff:(Harvard or HCO or Smithsonian or SAO)} \\
Positional searches & {\it pos(aff:``SAO'', 2)} \\
Search for citations & {\it citations(author:``Accomazzi, A'')} \\
Remove self-citations & {\it citations(author:``Accomazzi, A'') -author:``Accomazzi, A''} \\
Citations \& acknowledgments & {\it citations(author:``Accomazzi'') or ack:``Accomazzi''} \\
Review papers & {\it instructive(``star formation'')} \\
Popular papers & {\it trending(``star formation'')} \\
\noalign{\smallskip}
\tableline
\end{tabular}
}
\end{center}
\end{table}

\subsection{The User Interface}

The new ADS user interface has been completely revamped, and now follows the paradigm of modern search 
engines which provide a simple ``one-box'' input field for searching and multiple filters for
further refining.  This approach is quite different from
the traditional ADS fielded search interface, which today contains more than 100 elements 
(counting all input fields, checkboxes and radio buttons).
Usability testing has indicated that most users overall prefer the look of new interface when introduced to it,
although several power-users continue to find the traditional ADS Classic search form itself more
familiar and tend to resist the change.  One challenge we face is providing a friendly way
to introduce users to the new syntax, which includes learning about search fields, boolean
logic, and functional queries, although simple unfielded searches are supported as well.

The new interface takes full advantage of SOLR's native support for facets, which the ADS Labs
interface exposes as a set of filters on the left-hand bar (see Figure 1).  
This provides an intuitive and unobtrusive way for people to narrow the list of results to 
one's liking.
Of particular 
importance to librarians is the presence among these filters of the ADS bibliographic groups
and links to datasets.

\begin{figure}[!ht]
\plotone{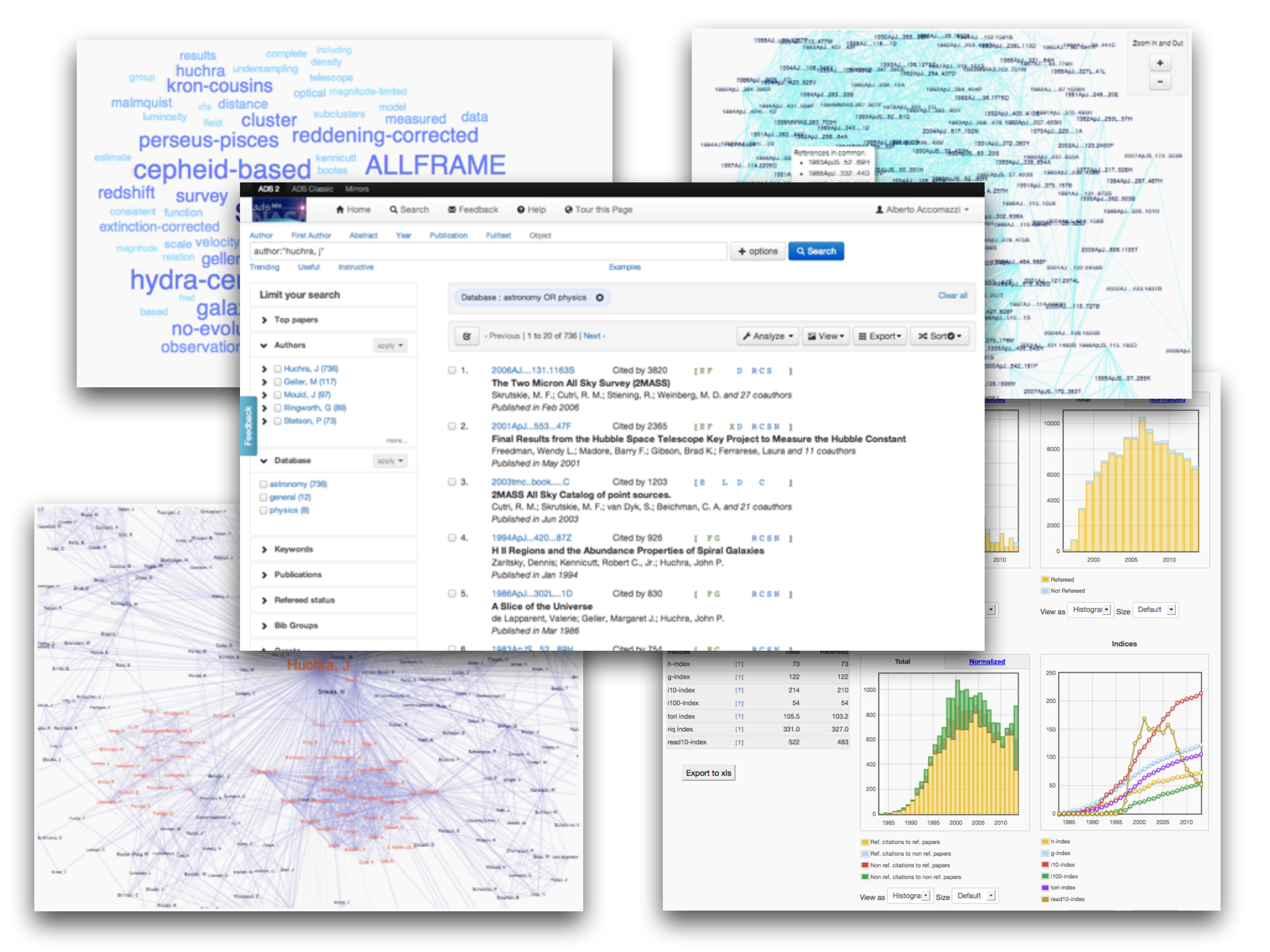}
\caption{The ADS Labs User Interface circa 2014.  
In the center of the figure is a list of results generated by an ADS query and 
surrounding it are visualizations generated from it.  
From top-left, in a clockwise order: word cloud, citation-based paper network,
metric summary, and co-authorship network.}
\end{figure}

The UI described here is currently evolving, and its underlying architecture is being updated
to take into account the best practices in web usability and responsive design.  
We are aware of the difficulty of providing a simple interface to a sophisticated search system
and a complex set of information resources, and are working with experts in the field of
visualizations, user-interfaces and web usability standards to help us find 
the right balance between functionality and simplicity.
We hope that the final result of our effort will be welcome by the community, yet are fully
aware of the natural aversion to changes requiring users to re-learn how to interact with
a system as popular as the ADS.

%%%%%%%%%%
\section{Enhanced Curation Support}

One of the goals behind the architecture of the new ADS platform is facilitating 
curation workflows of both ADS staff and collaborators.  One of the important uses
of ADS is the support of the typical activities of librarians and archivists who
make use of the ADS as both a discovery and curation platform.
In this section we highlight the features which we are providing to support
these activities.

\subsection{ADS Libraries}

The new ADS interface extends the notion of Private Libraries which were introduced in
ADS Classic over 10 years ago.  Libraries now allow users to bookmark, tag and
annotate records that may be of interest later. For example, one could create a library
containing a list of one's publications, or a set of references for a paper currently 
being written, or simply a set of interesting papers on a research topic. 
Libraries can be shared with one or more ADS users, and write permission can be granted
to such users by the owner if desired.  This is a good way to share a collection of 
items with one or more collaborators while still retaining ownership and curation 
responsibilities for the library.  Owners can even choose to make one or several of
their libraries publicly accessible, and then share the public URIs of those libraries
with interested parties.

The new ADS libraries interface also supports the notion of collaborative groups.  Groups consist
of sets of users who share a common set of libraries and who typically work as part of
a long-term collaboration requiring full access to a common library of papers.
For example, one could create groups for one's research collaborations, to share
materials with graduate students and post-docs, or to support a journal club. 
All members of a group can post items to the group's library, making it possible to
jointly contribute to the reading material in it.

Libraries in ADS Classic can be imported into the new system and provide a great
way for curators to jointly work on the maintenance of a bibliography, as the system
keeps provenance information associated with tags and annotations made to all
records in them.

\subsection{Bibliographic Groups}

The ADS provides a way for users to view curated collections of records related to 
a number of well-known astronomy projects and repositories.  These collections 
(commonly referred to as ``ADS bibliographic groups'') can be used as
search filters (``show me all papers on AGNs which have data from the Hubble Space Telescope'')
or as browsable lists of papers (``show me all papers in the ESO telescope bibliography'').
Bibliographic groups are a special case of ADS libraries, in that they are exposed 
by the ADS search engine to the general public and as such have a much higher visibility 
than any given individual user's library (assuming it has been made publicly accessible).
While the ADS Classic platform had a maximum number of bibliographic groups 
which could be defined in the system, the new ADS platform removes this limitation,
making it possible for us to support a much larger number of groups in the future.
One thing worth pointing out is that if you are already using the ADS libraries system to
maintain a bibliography, it is trivial to convert this to an ADS bibliographic group,
and expose it to the general public.

The primary goal of such bibliographies has always been to facilitate the discovery of data
through bibliographic sources, however the availability of paper-based metrics makes it possible
for archivists and program managers to assess the impact of an instrument, facility, funding
or observing program.  While this is an inevitable outcome of the bibliography curation 
process, we urge caution whenever impact analysis is performed using bibliographic data,
and urge people to read the ``fine print'' whenever using
and comparing paper-based metrics.
See the paper by \cite{2014arXiv1406.4542H} for a discussion of the metrics available from ADS.

The current list of bibliographic groups is found in Table 2.  It is worth pointing out that 
the lists of records which are part of a bibliographic group are created and maintained by curators
who are close ADS collaborators (but who do not work for ADS).  While there is now an emerging consensus
within the astronomical library community about what these bibliographies should contain, we should
point out that there is no single (or simple) set of criteria which is applied across all such
collections (nor could we consider enforcing one at this point).  To promote transparency about
the process behind their creation, the ADS help pages
\href{http://doc.adsabs.harvard.edu/abs_doc/help_pages/search.html\#Select_References_From_Group}{provide a short description of the bibliographic groups}.
While this is a good start, we are aware that a much more thorough description of the curation
process is desirable, and we will work with the community of Astronomy Librarians to 
improve the current situation.
Going forward, it is our intention to make it as clear as possible where the original data came from and
what the selection criteria behind each bibliography are.
If you are current curating a bibliography which may be of interest to the general Astronomy
community and would like to see it included in the ADS search interface, feel free to contact us.

\begin{table}[!ht]
\caption{The list of bibliographic groups currently in ADS.
As can be seen, most of them correspond to bibliographies related to
major astronomy archives, missions, or institutions.}
\smallskip
\begin{center}
{\small
\begin{tabular}{lllllll}
\tableline
\noalign{\smallskip}
ARI & CfA & CFHT & Chandra & ESO & Gemini & Herschel \\
HST & ISO & IUE & JCMT & Keck & Leiden & LPI \\
Magellan & NOAO & NRAO & ROSAT & SDO & SMA & Spitzer \\
Subaru & Swift & UKIRT & USNO & XMM \\
\noalign{\smallskip}
\tableline
\end{tabular}
}
\end{center}
\end{table}

\subsection{ORCID Integration}

High on the list of priorities for science librarians is promoting the adoption
of ORCID by the science community.  The use of ORCID, once fully implemented,
will provide a straightforward way for bibliographers to track not only staff
publications, but also datasets, software, and funding sources \citep{ORCID}.
ADS plans to provide support for ORCID in at least three different ways.
First, we are collecting ORCIDs provided to us with article metadata by
the publishers, and these will become searchable in the ADS interface shortly.
Second, we will provide a way for users to link an ADS user account with the appropriate
ORCID account.  Third, we will enable export of ADS records to ORCID, thus facilitating
the creation of one's bibliography inside the ORCID system.  Deeper integration
between the two systems will depend on available resources and feedback from
the community.

\subsection{Affiliation Search}

Affiliation searches are now supported by the new ADS search engine but 
the proper identification of papers via affiliation remains difficult due
to the lack of a controlled list of institutions, as discussed by \cite{Egret}.
To ameliorate the situation, we are in the process of mapping the affiliation fields currently
indexed in our database to a standard reference system maintained by
Ringgold, Inc.  This will allow users and curators to more easily identify
papers which have a particular affiliation specified in their metadata.
We expect that we will need help from the community to ensure that all the
relevant astronomy institutions are properly represented in the canonical
list that we will be using, and that their hierarchy is correct.
The paper by \cite{ADSAFF} describes this effort in greater detail.

\subsection{The ADS API}

Developers and curators who require regular, automated access to our search engine are encouraged 
to use the ADS Application Programming Interface (API).
API access requires registration so that we can keep track of usage and to ensure that we abide
by existing agreements with publishers.  Each API user is issued a personal developer key which
can then be used to access the search engine programmatically.
Our API currently exposes search as well as analytics capabilities: using the API one can
perform the same searches and retrieve the same metrics currently supported by the ADS Labs interface.
The functionality currently missing from the API (but planned to be rolled out later this year) is 
access to the ADS libraries system, the data behind the ADS visualization services 
and export of records in alternate formats.

It has been rewarding to notice the take-up in API usage since its launch in 2013.  As of
August 2014 there are over 100 people who signed up for access, despite the fact that there
is already a well-established API exposed by ADS Classic which has been in use for over a decade.
We hope that in due time we will migrate all API access to the new system so we can better
support a wide range of applications.
More information on the API is available at \url{https://github.com/adsabs/adsabs-dev-api}.

\section{Conclusions}

It is an exciting time to be working in the digital library landscape, where 
resources available online are still growing at an exponential pace.  
The challenge we face is finding the right stack of tools and workflows
to organize, curate, and expose these data.  The additional challenge for ADS
has been finding 
a way to innovate its platform while still maintaining the existing ADS ``classic''
system available and current.  This has forced us to turn this effort into a
long-term evolutionary process which allows us to incrementally transition
databases, web services, user interface, and curation workflows to minimize
disruption to our users and collaborators.

We are thankful for the contributions and feedback that the astronomy library community provides to 
ADS, and readily acknowledge them.  Librarians are natural partners in that
they understand our domain and appreciate the issues we struggle with every day,  
such as metadata quality, copyright permissions, digital resource management, 
preservation of research data and artifacts. 
Librarians are today on the front lines of efforts to preserve
data products and link them to bibliographies, software, funding sources, people, and institutions.
These are goals that we all share, and invite you to participate in the effort as our partner.  
If you want to see the fruits of your curation efforts widely shared, consider providing your
collections, annotations, and bibliographies to the ADS so that they can be integrated
into our platform for the benefit of the worldwide astronomy community.

\begin{acknowledgements}
This work has been supported by the NASA Astrophysics Data System project, funded by NASA grant NNX12AG54G.
\end{acknowledgements}

\end{document}